\documentclass[superscriptaddress,
twocolumn,
 amsmath,amssymb,
]{revtex4-2}

\usepackage{graphicx}  % for including images
\usepackage{microtype} % for typographical enhancements
\usepackage{amsmath}   % for equations and mathematics
\usepackage{hyperref}  % for hyperlinks
\usepackage[a4paper,top=4.2cm,bottom=4.2cm,left=3.5cm,right=3.5cm]{geometry}
\usepackage{menukeys}
\usepackage{pdfsync}
\usepackage{boldline}
\usepackage[version=3]{acro}
\usepackage[nolong,nosuper]{glossaries}  % Prevent loading longtable and supertabular
\setglossarystyle{list}  % Use list style
\usepackage{gensymb}
\usepackage{anysize}
\usepackage{epsfig}
\usepackage{cprotect}
\usepackage{verbatim}
\usepackage{bm}
\usepackage{etoolbox}
\usepackage{amssymb}
\usepackage{multirow}
\usepackage{yfonts}
\usepackage{xr-hyper}
\usepackage{mathrsfs}
\usepackage{enumitem}
\usepackage{empheq}
\usepackage[font=small,labelfont=bf]{caption}
\usepackage[title,toc,titletoc,page]{appendix}
\usepackage[capitalise]{cleveref}
\usepackage[most]{tcolorbox}
\usepackage{eurosym}
\usepackage{wrapfig}
\usepackage{scalerel}[2016/12/29]
\usepackage{leftidx}
\usepackage{tikz}
\usepackage{xcolor}
\usepackage{fancyvrb}

\usetikzlibrary{math} %needed tikz library
\crefname{enumi}{Case}{Cases}

\acsetup{
        make-links ,
        pages / display = first ,
        pages / fill    = {, },
        patch / floats  = false
}

\hypersetup{colorlinks=true,
        linkcolor=blue,
        anchorcolor=brown,
        citecolor=red,
        urlcolor=purple
}

\providecommand{\pder}[2]{\frac{\partial {#1}}{\partial {#2}}}

\providecommand{\der}[2]{\frac{\dint{#1}}{\dint{#2}}}
\providecommand{\dern}[3]{\frac{{\dint{}^#3} #1}{{\dint{#2}^#3}}}
\providecommand{\pder}[2]{\frac{\partial{#1}}{\partial{#2}}}

\providecommand{\bw}{\begin{widetext}}
        \providecommand{\ew}{\end{widetext}}

\providecommand{\newlinenn}{\\ \nn}

\definecolor{glossary-link-color}{RGB}{129, 19, 49}
\definecolor{revision-color}{RGB}{100,0,0}

\providecommand{\mwalleq}{{\mathord{%
                \tikz[baseline=-0.1ex]{

                    \tikzmath{\size = 0.03ex;  }

                    \draw[line width=1pt] (0,0) -- (\size,0);
                    \draw[line width=1pt, draw opacity = 0.15] (\size,0) -- (\size,\size);
                    \draw[line width=1pt] (\size,\size) -- (0,\size);
                    \draw[line width=1pt, draw opacity = 0.15] (0,\size) -- (0,0);
                }%
            }}}

\providecommand{\mtopeq}{{\mathord{%
                \tikz[baseline=-0.1ex]{

                    \tikzmath{\size = 0.03ex;  }

                    \draw[line width=1pt, draw opacity = 0.15] (0,0) -- (\size,0);
                    \draw[line width=1pt, draw opacity = 0.15] (\size,0) -- (\size,\size);
                    \draw[line width=1pt] (\size,\size) -- (0,\size);
                    \draw[line width=1pt, draw opacity = 0.15] (0,\size) -- (0,0);
                }%
            }}}
\providecommand{\mbottomeq}{{\mathord{%
                \tikz[baseline=-0.1ex]{

                    \tikzmath{\size = 0.03ex;  }

                    \draw[line width=1pt] (0,0) -- (\size,0);
                    \draw[line width=1pt, draw opacity = 0.15] (\size,0) -- (\size,\size);
                    \draw[line width=1pt, draw opacity = 0.15] (\size,\size) -- (0,\size);
                    \draw[line width=1pt, draw opacity = 0.15] (0,\size) -- (0,0);
                }%
            }}}

\providecommand{\myineq}{{\mathord{%
                \tikz[baseline=-0.1ex]{

                    \tikzmath{\size = 0.03ex;  }

                    \draw[line width=1pt, draw opacity = 0.15] (0,0) -- (\size,0);
                    \draw[line width=1pt, draw opacity = 0.15] (\size,0) -- (\size,\size);
                    \draw[line width=1pt, draw opacity = 0.15] (\size,\size) -- (0,\size);
                    \draw[line width=1pt] (0,\size) -- (0,0);
                }%
            }}}

\providecommand{\myinwalleq}{{\mathord{%
                \tikz[baseline=-0.1ex]{

                    \tikzmath{\size = 0.03ex;  }

                    \draw[line width=1pt] (0,0) -- (\size,0);
                    \draw[line width=1pt, draw opacity = 0.15] (\size,0) -- (\size,\size);
                    \draw[line width=1pt] (\size,\size) -- (0,\size);
                    \draw[line width=1pt] (0,\size) -- (0,0);
                }%
            }}}

\providecommand{\myouteq}{{\mathord{%
                \tikz[baseline=-0.1ex]{

                    \tikzmath{\size = 0.03ex;  }

                    \draw[line width=1pt, draw opacity = 0.15] (0,0) -- (\size,0);
                    \draw[line width=1pt] (\size,0) -- (\size,\size);
                    \draw[line width=1pt, draw opacity = 0.15] (\size,\size) -- (0,\size);
                    \draw[line width=1pt, draw opacity = 0.15] (0,\size) -- (0,0);
                }%
            }}}

\providecommand{\msquareeq}{{\mathord{%
                \tikz[baseline=-0.1ex]{

                    \tikzmath{\size = 0.03ex;  }

                    \draw[line width=1pt] (0,0) -- (\size,0);
                    \draw[line width=1pt] (\size,0) -- (\size,\size);
                    \draw[line width=1pt] (\size,\size) -- (0ex,\size);
                    \draw[line width=1pt] (0ex,\size) -- (0ex,0ex);
                }%
            }}}

\providecommand{\mcirceq}{{\mathord{%
                \tikz[baseline=-0.1ex]{
                    \draw[line width=1pt] (0,0) circle [radius=0.5ex];
                }%
            }}}

\providecommand{\mcircineq}{{\mathord{%
                \tikz[baseline=-0.5ex]{
                    \draw[line width=1pt] (0,0) circle [radius=0.3ex];
                    \draw[line width=1pt, draw opacity = 0.15] (0,0) circle [radius=.7ex];
                }%
            }}}

\providecommand{\mcircouteq}{{\mathord{%
                \tikz[baseline=-0.5ex]{
                    \draw[line width=1pt, draw opacity = 0.15] (0,0) circle [radius=0.3ex];
                    \draw[line width=1pt] (0,0) circle [radius=.65ex];
                }%
            }}}

\DeclareRobustCommand{\libsymbol}{%
    \raisebox{0.8ex}{%
        \scalebox{0.55}{
            \begin{tikzpicture}[scale=1, baseline={(current bounding box.center)}]
                \def\s{80}
                \def\A{0.1}
                \def\xmax{1.5}
                \def\scalefactorletters{1.5}
                \def\radtodeg{360.0/(2.0*pi)}
                \def\degtorad{1.0/(360.0/(2.0*pi))}

                \draw[color={rgb,255:red,0; green,0; blue,255}, opacity=0.8, line width=2pt] plot[domain=-0.2:1.2, samples=100]
                (\x*\xmax, {\A*sin(2*pi*\x*\xmax*\radtodeg)});

                \node at (0*\xmax, {\A*sin(2*pi* 0*\xmax * \radtodeg)})
                [rotate= atan(\A *2*pi  * cos(2*pi * 0*\xmax * \radtodeg))] {\scalebox{\scalefactorletters}{i}};
                \node at (0.25*\xmax, {\A*sin(2*pi* 0.25*\xmax * \radtodeg)})
                [rotate= atan(\A *2*pi  * cos(2*pi * 0.25*\xmax * \radtodeg))] {\scalebox{\scalefactorletters}{r}};
                \node at (0.5*\xmax, {\A*sin(2*pi* 0.5*\xmax * \radtodeg)})
                [rotate= atan(\A *2*pi  * cos(2*pi * 0.5*\xmax * \radtodeg))] {\scalebox{\scalefactorletters}{e}};
                \node at (0.75*\xmax, {\A*sin(2*pi* 0.75*\xmax * \radtodeg)})
                [rotate= atan(\A *2*pi  * cos(2*pi * 0.75*\xmax * \radtodeg))] {\scalebox{\scalefactorletters}{n}};
                \node at (1.0*\xmax, {\A*sin(2*pi* 1.0*\xmax * \radtodeg)})
                [rotate= atan(\A *2*pi  * cos(2*pi * 1.00*\xmax * \radtodeg))] {\scalebox{\scalefactorletters}{e}};
            \end{tikzpicture}%
        }%
        \vspace{-.5cm}
        \hspace{-0.2cm}
    }
}

\providecommand{\be}{\begin{equation}}
        \providecommand{\ee}{\end{equation}}
\providecommand{\bsp}{\begin{split}}
        \providecommand{\esp}{\end{split}}
\providecommand{\bea}{\begin{eqnarray}}
        \providecommand{\eea}{\end{eqnarray}}
\providecommand{\beas}{\begin{eqnarray*}}
        \providecommand{\eeas}{\end{eqnarray*}}

\providecommand{\myineqcap}{{\protect \myineq}}

\providecommand{\mcirceqcap}{{\protect \mcirceq}}

\providecommand{\mtopeqcap}{{\protect \mtopeq}}
\providecommand{\mbottomeqcap}{{\protect \mbottomeq}}

\providecommand{\testfunc}[1]{{\nu_{#1}}}
\providecommand{\mcircineqcap}{{\protect \mcircineq}}
\providecommand{\mcircouteqcap}{{\protect \mcircouteq}}
\providecommand{\om}{\Omega}

\providecommand{\pom}{{\partial \om}}
\providecommand{\pom}{{\partial \om}}
\providecommand{\pomsqeq}{\pom_\msquareeq}
\providecommand{\pomineq}{\pom_\myineqcap}

\providecommand{\pomin}{$\pomineq$}
\providecommand{\pomout}{$\pomouteq$}
\providecommand{\pomouteq}{\pom_\myouteqcap}

\providecommand{\pominwalleq}{\pom_\myinwalleqcap}
\providecommand{\pomcirceq}{{\pom_\mcirceqcap}}
\providecommand{\pomcircone}{$\pomcirceqone$}
\providecommand{\pomcirctwo}{$\pomcirceqtwo$}
\providecommand{\pomcirceqone}{{\pom_{\mcirceqcap 1}}}
\providecommand{\pomcirceqtwo}{{\pom_{\mcirceqcap 2}}}

\providecommand{\pomcircin}{$\pom_\mcircineqcap$}
\providecommand{\pomcircout}{$\pom_\mcircouteqcap$}
\providecommand{\pomcircineq}{\pom_\mcircineqcap}
\providecommand{\pomweq}{\pom_\mwalleqcap}

\providecommand{\mwalleqcap}{{\protect \mwalleq}}

\providecommand{\pomcircouteq}{\pom_\mcircouteqcap}

\providecommand{\pomtopeq}{\pom_\mtopeqcap}

\providecommand{\pombottomeq}{\pom_\mbottomeqcap}

\providecommand{\manifold}{\mathscr{M}}

\providecommand{\kap}{\kappa}
\providecommand{\nab}{\nabla}
\providecommand{\newt}{\text{N}}
\providecommand{\kb}{k_{\text B}}
\providecommand{\mic}{\mu \met}

\providecommand{\pas}{\text{Pa}}
\providecommand{\met}{\text{m}}
\providecommand{\newt}{\text{N}}

\providecommand{\kel}{K}
\providecommand{\bx}{{\boldsymbol{x}}}

\providecommand{\neucl}{\hat{N}}
\providecommand{\lapbel}{\text{\fontsize{7pt}{7pt}\selectfont{LB}}}
\providecommand{\nablb}{\nab_\lapbel}
\providecommand{\nm}{\text{nm}}

\providecommand{\cellsize}{l}

\providecommand{\Rtwo}{{\mathbb{R}}^2}

\providecommand{\myouteqcap}{{\protect \myouteq}}
\providecommand{\myinwalleqcap}{{\protect \myinwalleq}}

\providecommand{\baligned}{\begin{equation}\begin{aligned}}
            \providecommand{\ealigned}{\end{aligned}\end{equation}}
\providecommand{\nn}{\nonumber}
\providecommand{\libname}{{\protect\libsymbol}}

\providecommand{\normcurve}{\textswab{n}}

\providecommand{\vr}{\hat{x}}

\providecommand{\dint}[1]{\text{d} {#1}}

\ProvideDocumentCommand{\crefs}{m}{%
    \begingroup
    \crefname{equation}{}{}%
    \cref{#1}%
    \endgroup
}

\providecommand{\revision}[1]{{\color{black}#1}}

\DeclareAcronym{irene}{
short = {IRENE},
long = flu{I}d laye{R} finit{E}-{E}lement softwar{E},
}

\DeclareAcronym{sd}{
        short = {SD},
        short-plural-form = {SDs},
        long = small deformation,
}

\DeclareAcronym{guv}{
        short = {GUV},
        short-plural-form = {GUVs},
        long = giant unilamellar vesicle,
        long-plural-form = giant unilamellar vesicles,
}

\DeclareAcronym{ld}{
        short = {LD},
        short-plural-form = {LDs},
        long = large deformation,
}

\DeclareAcronym{br}{
        short = {BR},
        short-plural-form = {BRs},
        long =  bacteriorhodopsin ,
        long-plural-form =  bacteriorhodopsins,
}

\DeclareAcronym{fe}{
        short = {FE},
        short-plural-form = {FEs},
        long = finite element,
}
\DeclareAcronym{ode}{
        short = {ODE},
        short-plural-form = {ODEs},
        long = ordinary differential equation,
        long-plural-form =  ordinary differential equations,
}

\DeclareAcronym{ale}{
        short = {ALE},
        short-plural-form = {ALEs},
        long = arbitrary Lagrangian-Eulerian ,
}

\DeclareAcronym{prin}{
        short = {PI},
        short-plural-form = {PIs},
        long = {protein inclusion},
        long-plural-form =  {protein inclusions},
}

\DeclareAcronym{lb}{
        short = {LB},
        long = \ac{lb},
}

\DeclareAcronym{vp}{
        short = {VP},
        short-plural-form = {VPs},
        long = variational problem,
        long-plural-form = variational problems,
}

\DeclareAcronym{cn}{
        short = {CN},
        long = Crank Nicolson,
}
\DeclareAcronym{ipcs}{
        short = {IPCS},
        long = incremental pressure correction scheme,
}

\DeclareAcronym{lhs}{
        short = {LHS},
        short-plural-form = {LHSs},
        long = left-hand side,
        long-plural-form = left-hand sides,
}
\DeclareAcronym{rhs}{
        short = {RHS},
        short-plural-form = {RHSs},
        long = right-hand side,
        long-plural-form = right-hand sides,
}

\DeclareAcronym{bc}{
        short = {BC},
        short-plural-form = {BCs},
        long = {boundary condition},
        long-plural-form = {boundary conditions}
}

\DeclareAcronym{bvp}{
        short = {BVP},
        short-plural-form = {BVPs},
        long = {boundary-value problem},
        long-plural-form = {boundary-value problems}
}

\DeclareAcronym{ns}{
        short = {NS},
        long = {Navier-Stokes},
}

\DeclareAcronym{fem}{
        short = {FEM},
        short-plural-form = {FEMs},
        long = {\ac{fe} method},
        long-plural-form = {finite-element methods}
}

\DeclareAcronym{pde}{
        short = {PDE},
        short-plural-form = {PDEs},
        long = {partial differential equation},
        long-plural-form = {partial differential equations}
}
\DeclareAcronym{fenics}{
        short = {FEniCS},
        long = {finite element computational software},
}

\providecommand{\vthreshold}{v_\ast}
\providecommand{\papertitle}{Interaction between cell membranes and  protein inclusions in the large-deformation regime}

\providecommand{\lambdasd}{\lambda_\text{SD}}
\providecommand{\etathreed}{\eta_\text{3D}}
\providecommand{\scrivenlovenum}{S_\text{L}}

\newglossaryentry{element}{
        name={element},
        description={an atomic part of a mesh \cite{zienkiewiczFiniteElementMethod2013,liuLectureNotesIntroduction1997}},
        plural={elements},
}

\newglossaryentry{r}{
        name={$r$},
        description={radius of circular obstacle in a mesh},
}

\newglossaryentry{manifold}{
        name={\ensuremath{\manifold}},
        description={differential manifold \cite{marchiafavaAppuntiDiGeometria2005}},
}

\newglossaryentry{h_omega}{
        name={\ensuremath{h}},
        description={height of a rectangle which defines \gls{omega_z}},
}

\newglossaryentry{L_omega}{
        name={\ensuremath{L}},
        description={length of a rectangle which defines \gls{omega_z}},
}

\newglossaryentry{rho}{
        name={\ensuremath{\rho}},
        description={density \cite{landauFluidMechanics1987}},
}

\newglossaryentry{eta}{
        name={\ensuremath{\eta}},
        description={two-dimensional viscosity \cite{landauFluidMechanics1987}},
}

\newglossaryentry{c}{
        name={$\bf{c}$},
        description={center of the circular obstacle in a mesh},
}

\newglossaryentry{vr}{
        name={$\vr$},
        description={radial direction: $\vr^i = \frac{x^i}{|\bf x|}$},
        plural={elements},
}
\newglossaryentry{omega}{
        name={\ensuremath{\om}},
        description={subset of $\Rtwo$ over which the coordinates of \gls{manifold} are defined \cite{evansPartialDifferentialEquations2010}},
}

\newglossaryentry{h}{
        name={$h$},
        description={pull-back of the metric $g$ on a curve \gls{curve} in \gls{manifold} \cite{marchiafavaAppuntiDiGeometria2005,reallGeneralRelativity}},
}

\newglossaryentry{pomega}{
        name={\ensuremath{\pom}},
        description={boundary of  \gls{omega}, see \cref{fig-geometry,definition-pom}},
}

\newglossaryentry{curve}{
        name={\ensuremath{\gamma}},
        description={a curve in \gls{manifold}},
}

\newglossaryentry{normalcurve}{
        name={\ensuremath{\normcurve}},
        description={vector normal to a curve \gls{curve} in \gls{manifold}; this vector belongs to the tangent bundle of \gls{manifold} \cite{marchiafavaAppuntiDiGeometria2005}},
}

\newglossaryentry{cellsize}{
        name={\ensuremath{\cellsize}},
        description={mesh cell size: the smallest cell diameter, across all cells in the mesh},
}

\newglossaryentry{pomegasq}{
        name={\ensuremath{\protect\pomsqeq} },
        description={rectangular boundary of  \gls{omega}, see \cref{definition-pomsq}},
}
\newglossaryentry{pomegaci}{
        name={\ensuremath{\protect\pomcirceq} },
        description={circular boundary of  \gls{omega}},
}

\newglossaryentry{pomegain}{
        name={\ensuremath{\protect  \pomineq} },
        description={boundary of  \gls{omega} located at the left edge of the rectangle} ,
}

\newglossaryentry{pomegaout}{
        name={\ensuremath{\protect  \pomouteq} },
        description={same as \gls{pomegain}, for the right edge of the rectangle} ,
}

\newglossaryentry{pomegainwall}{
        name={\ensuremath{\protect  \pominwalleq} },
        description={\gls{pomegain} $\cup$ \gls{pomegaw}} ,
}

\newglossaryentry{pomegatop}{
        name={\ensuremath{\protect  \pomtopeq} },
        description={boundary of  \gls{omega} located at the top edge of the rectangle} ,
}
\newglossaryentry{pomegabottom}{
        name={\ensuremath{\protect  \pombottomeq} },
        description={boundary of  \gls{omega} located at the bottom edge of the rectangle} ,
}

\newglossaryentry{pomegaw}{
        name={\ensuremath{\protect  \pomweq} },
        description={same as \gls{pomegain}, for the top and bottom edges of the rectangle, see \cref{definition-pomwall}} ,
}

\newglossaryentry{pomegacircin}{
        name={\ensuremath{\protect\pomcircineq} },
        description={inner circular boundary of \gls{omega}},
}

\newglossaryentry{pomegacircout}{
        name={\ensuremath{\protect\pomcircouteq} },
        description={same as \gls{pomegacircin}, for the outer circular boundary},
}

\newglossaryentry{nab}{
        name={$\nab$},
        description={covariant derivative \cite{marchiafavaAppuntiDiGeometria2005}},
        plural={covariant derivatives},
}

\newglossaryentry{g}{
        name={$g$},
        description={metric tensor \cite{marchiafavaAppuntiDiGeometria2005}},
        plural={metric tensors},
}

\newglossaryentry{b}{
        name={$b$},
        description={second fundamental form \cite{marchiafavaAppuntiDiGeometria2005}},
        plural={second fundamental forms},
}

\newglossaryentry{H}{
        name={$H$},
        description={mean curvature \cite{desernoNotesDifferentialGeometry2004}},
        plural={mean curvatures},
}
\newglossaryentry{gausscurv}{
        name={$K$},
        description={Gaussian curvature \cite{desernoNotesDifferentialGeometry2004}},
        plural={mean curvatures},
}

\newglossaryentry{nablalb}{
        name={$\nab_{\lapbel}$},
        description={Laplace-Beltrami operator \cite{arroyoRelaxationDynamicsFluid2009i}},
        plural={Laplace-Beltrami operators},
}

\newglossaryentry{neucl}{
        name={$\neucl$},
        description={unit vector in the three-dimensional Euclidean space, normal to \gls{manifold}, see \cref{fig-geometry}},
}
\newglossaryentry{ntan}{
        name={$n$},
        description={unit vector in the tangent bundle of \gls{manifold} and normal to a curve in \gls{manifold}, see \cref{fig-geometry}},
}

\newglossaryentry{bx}{
        name={$\bx$},
        description={coordinates on \gls{manifold}},
}

\newglossaryentry{kappa}{
        name={$\kap$},
        description={bending rigidity \cite{derenyiFormationInteractionMembrane2002}},
}

\newglossaryentry{v}{
        name={$v$},
        description={tangential velocity, it is a vector field the tangent bundle of \gls{manifold}},
}

\newglossaryentry{w}{
        name={$w$},
        description={normal velocity, it is a scalar on  \gls{manifold}},
}

\newglossaryentry{omega_z}{
        name={$\omega$},
        description={gradient of \gls{z}, it is a one-form on \gls{manifold}},
}

\newglossaryentry{omega_r}{
        name={$\omega_r$},
        description={radial component of \gls{omega_z} in polar coordinates},
}

\newglossaryentry{mu}{
        name={$\mu$},
        description={auxiliary variable which equals the mean curvature \gls{H}, see \cref{eq-def-mu}. It is a scalar on  \gls{manifold}},
}

\newglossaryentry{sigma}{
        name={$\sigma$},
        description={surface tension \cite{derenyiFormationInteractionMembrane2002}, it is a scalar on  \gls{manifold}},
}
\newglossaryentry{z}{
        name={$z$},
        description={fluid shape profile, it is a scalar on  \gls{manifold}},
}

\newglossaryentry{testf}{
        name={\ensuremath{\protect\testfunc{}}},
        description={test function in \ac{fe} methods \cite{zienkiewiczFiniteElementMethod2013}. In \libname, it is  denoted by the suffix of its related function, e.g., the test function related to $z$ is $\testfunc{z}$},
}

\newglossaryentry{fenics}{
        name={\ac*{fenics}},
        description={\acl{fenics} \cite{loggAutomatedSolutionDifferential2012}, on which \libname is built.},
}

\newglossaryentry{eps}{
        name={$\epsilon_{ij}$},
        description={Levi-Civita antisymmetric symbol \cite{marchiafavaAppuntiDiGeometria2005}},
        plural={elements},
}

\crefname{enumi}{Point}{Points}
\Crefname{enumi}{Point}{Points}

\begin{document}

\setlength\intextsep{0pt}
\setlength{\parskip}{5pt plus 0pt minus 0pt}

\preprint{APS/123-QED}

\title{\papertitle}

\author{Gaetano Ferraro}
\affiliation{Institut Curie, PSL Research University, Paris, France}
\affiliation{CNRS UMR168, 11 rue Pierre et Marie Curie, 75005, Paris, France}
\affiliation{Polytechnic University of Turin, Corso Castelfidardo 39, 10129 Turin, Italy}

\author{Michele Castellana}
\thanks{Corresponding author: \href{mailto:michele.castellana@curie.fr}{michele.castellana@curie.fr}}
\affiliation{Institut Curie, PSL Research University, Paris, France}
\affiliation{CNRS UMR168, 11 rue Pierre et Marie Curie, 75005, Paris, France}

\date{\today}
\begin{abstract}
  Biological membranes are dynamic surfaces whose shape and function are critically influenced by \acp{prin}. While membrane deformations induced by \acp{prin} have been extensively studied in the small-deformation regime, a  variety of processes  \revision{involve} strong membrane deformations.  We investigate the interaction between lipid membranes and \ac{prin}s in the \ac{ld} regime, with the finite-element method. We  develop an approximate  analytical solution that captures key features of the \ac{ld} regime. We show that the force exerted by the membrane on a \ac{prin} displays non-monotonic behavior with respect to the \ac{prin} vertical displacement. The qualitative features of this force appear to be independent of the protein geometry. For two interacting \ac{prin}s, the membrane-mediated potential exhibits sub-power-law decay with inter-protein distance, reflecting the complex nature of the elastic medium. The interaction potential shows that conical \ac{prin}s with identical and opposite orientations repel and attract, respectively, confirming the analogy between \ac{prin} orientation and electric charge, in the \ac{ld} regime. In the presence of membrane flows, we identify a characteristic velocity that separates two regimes in which bending rigidity and viscous effects dominate, respectively, implying the onset of flow-induced deformations above such velocity threshold. Overall, our results provide quantitative predictions for membrane-protein systems in biologically relevant scenarios involving \acp{ld}, with implications for protein sorting, clustering, and membrane trafficking.
\end{abstract}

\maketitle

\acresetall
\section{Introduction} \label{sec_intro}
Biological membranes are dynamic and deformable surfaces whose shape and function are critically influenced by the presence of \acp{prin} \cite{alberts2022molecular,bethaniSpatialOrganizationTransmembrane2010}. Such \acp{prin} often impose geometric constraints, e.g., contact angles between the protein and the membrane, or membrane displacement, that drive local membrane deformations \cite{mannevilleActivityTransmembraneProteins1999} and may govern membrane-mediated interactions \cite{koltoverMembraneMediatedAttraction1999}. Understanding how these deformations arise and interact is central to elucidate key biological processes, such as cell-shape regulation and protein sorting \cite{mcmahonMembraneCurvatureMechanisms2005}.

Membrane deformations induced by \acp{prin} have been the subject of numerous investigations \cite{marshProteinModulationLipids2008}.  From the modelling standpoint, they may be described by the shape equation \cite{julicherShapeEquationsAxisymmetric1994,derenyiFormationInteractionMembrane2002,zhong-canBendingEnergyVesicle1989}, whose variational formulation stems from the free energy originally proposed by Helfrich \cite{helfrichElasticPropertiesLipid1973}. Studies on such membrane deformations in the literature focus on the \ac{sd} regime \cite{goulianLongRangeForcesHeterogeneous1993,weiklInteractionConicalMembrane1998,dommersnesLongrangeElasticForces1998,kimCurvatureMediatedInteractionsMembrane1998}, and rely on  perturbative expansions about the flat-membrane configurations.

However, a wide variety of physical situations may fall beyond the \ac{sd} scenario. For instance, a \ac{prin} moving at large enough velocities may induce a strong membrane invagination. Also, multiple \acp{prin} positioned close enough to each other, may induce strong membrane deformations and gradients which cannot be described as \acp{sd}.

In this work, we explore the interaction between a lipid membrane and a \ac{prin}, in the \ac{ld} regime, leveraging the \ac{fe} method to solve the problem numerically. We then focus on the  features resulting from this solution, such as the membrane shape, flows,  the membrane-mediated interaction between two \acp{prin}, and others, and discuss their physical interpretation.

The paper is structured as follows. In \cref{sec:state_of_the_art} we discuss the state of the art on membrane-mediated forces. \Cref{sec_results} contains our results in the \ac{ld} regime. In particular, in \cref{sec_ss_no_flow} we focus on the steady state in the absence of flows. After discussing the limitations of the \ac{sd} solution in \cref{sec_linearized_equation}, in \cref{sec:nonlinear_approximation} we present an approximate,  analytical solution for the \ac{ld} regime. In \cref{sec_forces} we discuss the force exerted by the membrane on the \ac{prin}, and in \cref{sec_two_proteins} the membrane-mediated interaction between a pair of \acp{prin}. In \cref{sec_ss_flow} we study the steady state in the presence of flows by focusing on the effect of inflow velocity on membrane shape. Finally, \cref{sec:disc} is devoted to the discussion and interpretation of the results.

\section{State of the art}\label{sec:state_of_the_art}

\revision{Previous studies have investigated how the presence of \acp{prin} influences membrane shape and curvature. Among these, several works focused on the \ac{sd} regime \cite{dommersnesNbodyStudyAnisotropic1999,fournierCouplingMembraneTiltdifference1998, rangamaniProteinInducedMembraneCurvature2014}, employing mean-field models in which only the protein density is considered, with an additional coupling term between protein density and membrane curvature included in the free energy. Other studies have addressed the \ac{ld} regime \cite{powersFluidmembraneTethersMinimal2002, derenyiFormationInteractionMembrane2002}, analyzing both membrane shape and interaction forces; however, they model proteins as point-like inclusions, thereby neglecting finite-size effects and geometric constraints imposed by the proteins, such as the contact angle at their boundaries.}\\
On top of numerous experimental studies---see for example \cite{koltoverMembraneMediatedAttraction1999,mannevilleActivityTransmembraneProteins1999,Bassereau_2018}---membrane-mediated forces between proteins have been subject of interest and research in the past few decades from the theoretical standpoint. However, the existing theoretical studies are restricted to \acp{sd}---see below.
In the pioneering study of Goulian et al., a power-law interaction between protein inclusions has been found in the  \ac{sd} limit  \cite{goulianLongRangeForcesHeterogeneous1993}, and later on studied for conical inclusions, in the \ac{sd} regime, as a function of lateral membrane tension  \cite{weiklInteractionConicalMembrane1998} and of the inclusion contact angle \cite{dommersnesLongrangeElasticForces1998}.
Studies with more than two inclusions showed, for \acp{sd}, the presence of non-pairwise forces between the inclusions, which allow for the formation of  membrane-bound protein aggregates  \cite{kimCurvatureMediatedInteractionsMembrane1998}.
In addition, the power-law dependence on inclusion symmetries was studied in the \ac{sd} limit \cite{marchenkoElasticInteractionPoint2002}. The nature of this interaction was then studied, in the \ac{sd} regime, as function of the `hardness' of the protein-protein interaction potential \cite{bartoloElasticInteractionHard2003}.
A geometrical approach has been recently proposed and used to derive force-distance relations in the \ac{sd} regime  \cite{muellerInterfaceMediatedInteractions2005}. Expressions for the inter-particle forces, not restricted to \acp{sd}, have been derived by leveraging the system symmetries \cite{muellerGeometrySurfaceMediated2005}, but they lack quantitative predictive power in the \ac{ld} regime due to the absence of a full numerical solution.

Regarding the effect of membrane flows on membrane shape, Arroyo and DeSimone \cite{arroyoRelaxationDynamicsFluid2009i} showed that membrane viscosity plays a crucial role in the relaxation dynamics of fluid membranes, highlighting the importance of viscous effects in dynamical models.
\revision{In \cite{sahuGeometryDynamicsLipid2020,tchoufagAbsoluteVsConvective2022},  membrane instabilities have been studied in simple geometries, showing that viscous effects can be characterized by a dimensionless number governing the influence of membrane viscosity on shape.}
\revision{In \cite{danielsCurvatureCorrectionMobility2016}  the effect of non-zero membrane curvature on the diffusion coefficient of \acp{prin} has been evaluated analytically, extending a result previously obtained in \cite{quemeneurShapeMattersProtein2014}, where  the effect of membrane deformation due to \acp{prin} in static condition had been studied. In both cases, the curvature-generation mechanism  has been considered as separated from  membrane flow.}
\revision{In \cite{morrisMobilityMeasurementsProbe2015}, the effect of membrane tension on the diffusion coefficient of \acp{prin} that locally deform the membrane by imposing a finite angle is analyzed analytically within a perturbative framework, valid in the limit of small imposed angles.}
Finally, in \cite{mahapatraTransportPhenomenaFluid2020}, coupled mean-field equations describing the time evolution of protein density and membrane shape were derived, accounting for the interplay between curvature elasticity and transport processes.

\revision{Regarding previous numerical studies, several works \cite{sauerStabilizedFiniteElement2017,barrettParametricFiniteElement2020,krause2023surface, contriFiniteElementFramework2025,sahuArbitraryLagrangianEulerian2024} have proposed \ac{fe} methods to model fluid membranes---see \cite{worthmullerIRENEFluIdLayeR2025} for  a detailed overview on  the features and limitations of such \ac{fe} studies.}

\section{Results}\label{sec_results}

% steady state with no flows
\subsection{Steady state with no flows}
\label{sec_ss_no_flow}

In this Section, we will present the results for a biological membrane at steady state with no flows, in a given surface-tension gradient. Because the presence of surface-tension gradients would imply the presence of flows, here we will consider constant surface-tension profiles, where the value of the surface tension is given by experimental data \cite{antonnyMembraneFissionDynamin2016}.

\label{sec:comments_on_the_no_flow_equation}
We describe the membrane shape by using the Monge gauge \cite{hsiungFirstCourseDifferential1981,desernoNotesDifferentialGeometry2004}: The membrane surface is a two-dimensional manifold embedded in three dimensions, described by a function $z(\bx)$, where $\bx = (x^1, x^2)$ are the manifold coordinates, which are defined in a two-dimensional domain $\om$, see for example \cref{fig:BCs-scheme_radial_symmetry,fig:BCs-scheme}. Given the tangent vectors to the coordinate lines ${\bm e}_1 \equiv (1, 0, \partial_1 z)$, ${\bm e}_2 \equiv (0, 1, \partial_2 z)$, the metric tensor is $g_{ij} = {\bm e}_i \cdot {\bm e}_j$ \cite{marchiafavaAppuntiDiGeometria2005}.  The membrane shape is determined by the  shape equation \cite{julicherShapeEquationsAxisymmetric1994,derenyiFormationInteractionMembrane2002,zhong-canBendingEnergyVesicle1989}
\begin{equation}
    \nabla^i \nabla_i H + 2H(H^2 - K) - H \frac{\sigma}{\kappa}=0.
    \label{eq:stationary_equation_general_form}
\end{equation}
In \cref{eq:stationary_equation_general_form}, $\nab$ is the covariant derivative associated to the Levi-Civita connection induced by $g$, $H$ and $K$  the mean and Gaussian curvature, see \cite{desernoNotesDifferentialGeometry2004,worthmullerIRENEFluIdLayeR2025} for details. Also, $\kap$ is the bending rigidity and $\sigma$ the surface tension \cite{desernoNotesDifferentialGeometry2004}.

\revision{The nondimensionalization of \Cref{eq:stationary_equation_general_form} leads to the appearance of the F\"oppl-von K\'arm\'an number \cite{powersFluidmembraneTethersMinimal2002,sahuGeometryDynamicsLipid2020}, defined as $\Gamma = \frac{\ell^2 \sigma}{\kappa}$. It characterizes the ratio between tension and bending effects and introduces a crossover length separating stretching-dominated regimes from bending-dominated ones, given by}
\begin{equation}
    \ell \equiv \sqrt{\frac{\kappa}{\sigma}}.
\end{equation}
For lengths smaller and larger than $\ell$, the bending rigidity and the surface tension  dominate the physical behavior of the system, respectively \cite{al-izziShearDrivenInstabilitiesMembrane2020}.
Given the  parameter values of   \cref{tab:physical_parameters}, $\ell$ is around $20 \,r_0$, where $r_0$ is the radius of the \ac{prin}  \cite{albertsMolecularBiologyCell2007,aivaliotisMolecularSizeDetermination2003}.\\

\begin{table}
    \centering
    \begin{tabular}{@{}lll@{}}
        \toprule
        \textbf{Parameter} & \textbf{Description} & \textbf{Value}                \\
        \toprule
        $r_0$              & Protein radius       & $10\,\nm$                     \\
        $\eta$             & Viscosity            & $10^{-8}\,\pas\, \met\, \sec$ \\
        $\sigma_0$         & Surface tension      & $10^{-6}\,\newt / \met$       \\
        $\kappa$           & Bending rigidity     & $10\ \kb T$                   \\
        $T$                & Temperature          & $300\,\kel$                   \\                            \\
        \toprule
    \end{tabular}
    \caption{Parameters for a lipidic cell membrane
        \cite{hormelMeasuringLipidMembrane2014,brochard-wyartHydrodynamicNarrowingTubes2006,morlotMembraneShapeEdge2012,quemeneurShapeMattersProtein2014,kozlovMembraneTensionMembrane2015,gauthierMechanicalFeedbackMembrane2012}.}
    \label{tab:physical_parameters}
\end{table}

\subsubsection{Linearized equation}\label{sec_linearized_equation}

\Cref{eq:stationary_equation_general_form} can be simplified in the case of a circular inclusion. Indeed, if the membrane shape is radially symmetric, \cref{eq:stationary_equation_general_form} reduces to an \ac{ode}. Moreover, \cref{eq:stationary_equation_general_form} can be linearized for small values of
\be
\omega \equiv   \pder{z}{r},
\ee
and it reduces to \cite{benderAdvancedMathematicalMethods1999}
\begin{align}\label{eq:stat_eq_linearized}
    \left(1- \frac{r^2}{\ell^2}\right)\omega-r\left(1+\frac{r^2}{\ell^2}\right)\der{\omega}{r}+2r^2\dern{\omega}{r}{2}+\newlinenn
    r^3 \dern{\omega}{r}{3} & = 0.
\end{align}

\begin{figure}
    \centering
    \includegraphics[width=1.1\columnwidth]{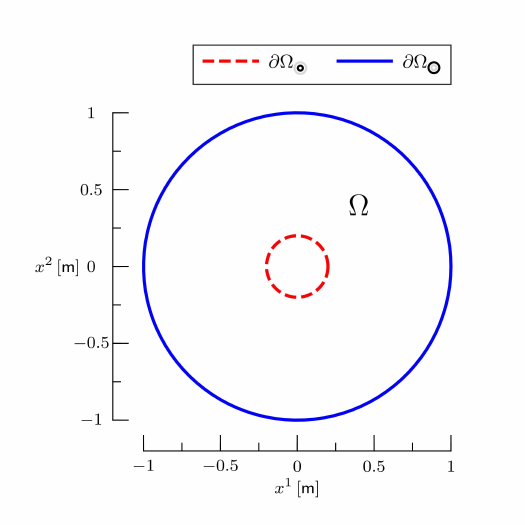}
    \caption{Sketch of the domain $\om$ and its boundaries for the radially symmetric steady state of a membrane and \acl{prin}, in the absence of flows.   The center of \pomcircin{} (red dashed circle) is located at $x^1=x^2=0$, and its radius is denoted by  $r_0$.  The center of \pomcircout{} (blue solid circle) is located in the same position, and its radius is denoted by  $R$. }
    \label{fig:BCs-scheme_radial_symmetry}
\end{figure}

\begin{figure*}[t]
    \centering
    \includegraphics[width=0.9\textwidth]{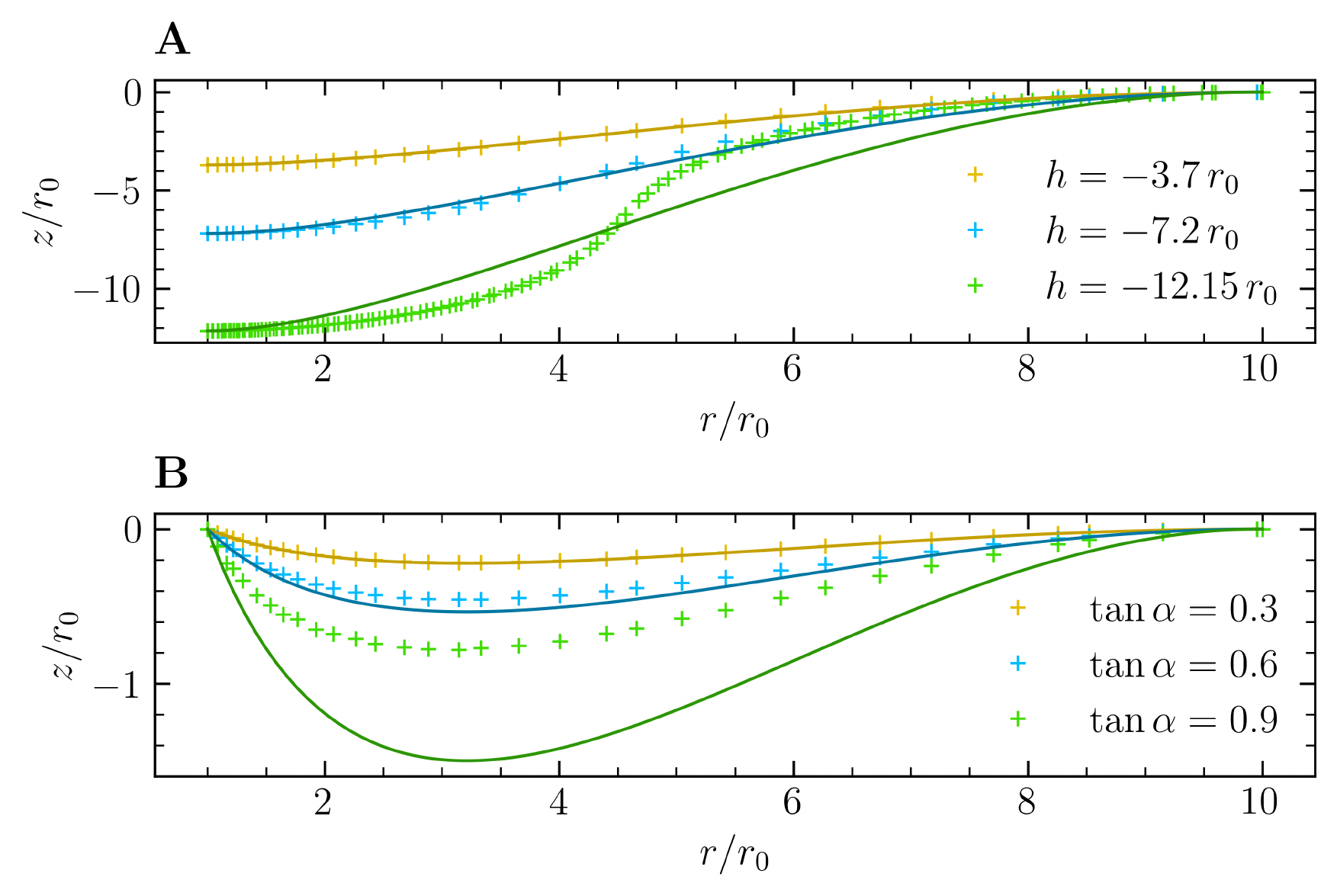}
    \caption{
        Comparison between analytical and numerical membrane shapes obtained with \ac{irene} \cite{worthmullerIRENEFluIdLayeR2025}.
        \textbf{A}) Comparison for fixed contact angle and different values of the protein vertical displacement $h$, with \aclp{bc} \crefs{linearized_BC_1,linearized_BC_2,linearized_BC_3,linearized_BC_4}.
        \textbf{B}) Same as \textbf{A}, with fixed protein displacement and different values of the contact angle
        $\alpha$.
    }
    \label{fig:comparison-sections}
\end{figure*}

\Cref{eq:stat_eq_linearized} has been extensively studied in the literature. However, its validity is limited to certain conditions:
In the following, we assess these limitations by comparing its predictions with numerical solutions.
To achieve this, we consider the \acp{bc}, see \cref{fig:BCs-scheme_radial_symmetry}:
\begin{align}
    z            & =       h_0      & \text{ on } \pomcircineq  & \, (r=r_0),     \label{linearized_BC_1}  \\
    z            & =         0      & \text{ on } \pomcircouteq & \, (r=R) ,       \label{linearized_BC_2} \\
    n^i \nab_i z & =  - \tan \alpha & \text{ on } \pomcircineq  & \, (r=r_0), \label{linearized_BC_3}      \\
    n^i \nab_i z & =   0            & \text{ on } \pomcircouteq & \, (r=R).\label{linearized_BC_4}
\end{align}
where $n^i$ is the unit normal, which lies in the membrane tangent bundle and points outside the membrane manifold \cite{worthmullerIRENEFluIdLayeR2025}.
\Cref{linearized_BC_1,linearized_BC_2,linearized_BC_3,linearized_BC_4} correspond to a membrane pinned on both boundaries \pomcircin{} and \pomcircout{}.
\revision{\Cref{linearized_BC_1,linearized_BC_3} imply a non-zero force and torque exerted by the membrane on the \ac{prin}, which are balanced by an external force pulling the protein and by the protein's internal elasticity, respectively. In the following, \Cref{linearized_BC_3} is enforced exactly by assuming the protein to be infinitely rigid. In reality, finite rigidity would lead to protein deformation, and the actual contact angle would result from a balance between membrane and protein torques \cite{danielsCurvatureCorrectionMobility2016}.}
\revision{\Cref{linearized_BC_2} arise from the presence of a pinning mechanism, such as the actin cortex, which constrains the membrane. Moreover, \Cref{linearized_BC_4} implies the existence of an applied torque, again associated with the pinning constraint.}
To assess the regime of validity of the linearized equation,  we compared membrane profiles from \ac{fe} solutions with the solution of \cref{eq:stat_eq_linearized}.  \Cref{fig:comparison-sections} shows that the linear solution remains accurate for imposed contact angles of up to approximately \(\tan \alpha \simeq 0.4\). For larger values of the contact angle, the linear approximation fails to capture the correct shape of the membrane and full non-linear equation \crefs{eq:stationary_equation_general_form} must be solved.

\begin{figure*}[t]
    \centering
    \includegraphics[width=1\textwidth]{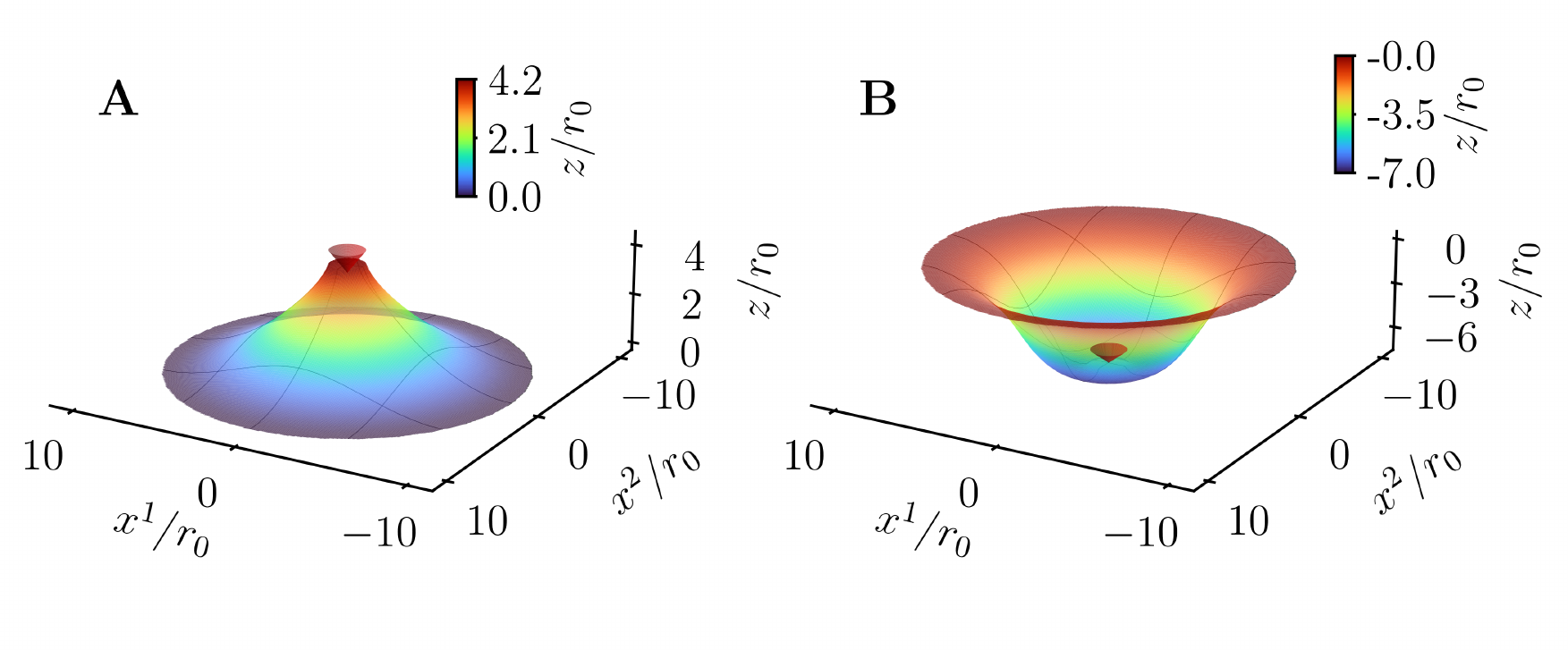}
    \caption{
        Membrane shapes in the large-deformation regime, obtained with the \acf{irene} \cite{worthmullerIRENEFluIdLayeR2025}.
        \textbf{A}) Membrane deformation induced by a positive vertical displacement of the protein and positive contact angle,  $h_0 = 4.2 \, r_0$ and  $\tan \alpha = 0.5$, cf. \cref{fig:BCs-scheme_radial_symmetry}.
        \textbf{B}) Membrane deformation induced by a negative vertical displacement of the protein and  positive contact angle angle, $h_0 = -6.5\, r_0$,  $\tan \alpha = 0.5$, cf. \cref{fig:BCs-scheme_radial_symmetry}.
        \revision{The \acp{prin} are depicted as red cones.}
        A convergence analysis of such type of radially symmetric solution with respect to the finite-element mesh resolution is reported in \cite{worthmullerIRENEFluIdLayeR2025}.
    }
    \label{fig:non_linear_regimes}
\end{figure*}

In what follows, we will focus on this regime, which is characterized by the presence of large gradients $\omega \gg 1$.
The \ac{prin} displacement and contact angle may have the same or opposite sign, leading to distinct deformation responses; these two situations are presented in \cref{fig:non_linear_regimes}.

\subsubsection{Solution with zero mean curvature} \label{sec:nonlinear_approximation}

Analytical insights about the \ac{ld} regime for the membrane can be obtained considering the following approximate solution.

In order to simplify the equation, it is convenient to introduce the quantity $\psi$, by the relation
\be\label{eq_sub_omega}
\omega = \frac{\psi}{\sqrt{1-\psi^2}}.
\ee
In \cref{appendix_zero_curvature}, we show all geometrical quantities expressed in terms of $\psi$ and its derivatives. Substituting \cref{eq_geo_sub_omega_psi} in \cref{eq:stationary_equation_general_form}, we obtain:
\begin{align}
    \label{eq_shape_1}
    (1-\psi^2)\frac{\partial^2 H}{\partial r^2} + \left(\frac{1-\psi^2}{r}-\psi \frac{\partial \psi}{\partial r}\right)\frac{\partial H}{\partial r}+ \nonumber \\
    \left(\frac{2\psi}{ r} \frac{\partial \psi}{\partial r} + \frac{1}{\ell}-2H^2\right)H=0 ,                                                                   \\
    \label{eq_shape_2}
    H = \frac{\partial \psi}{\partial r} + \frac{\psi}{r}.
\end{align}

\Cref{eq_shape_2} admits a  solution with zero mean curvature, of the form
\be\label{eq_zero_curvature}
\psi = \frac{C_1}{r}.
\ee
We observe that the two terms in the \acl{rhs} of \cref{eq_shape_2} are, respectively, the radial and angular curvature \cite{marchiafavaAppuntiDiGeometria2005}. As a result,  solution \crefs{eq_zero_curvature} does not imply that $z$ is linear in $r$, i.e., that the radial curvature vanishes, but that the algebraic sum of radial and angular curvature is zero.

\begin{figure}[t]
    \centering
    \includegraphics[width=\columnwidth]{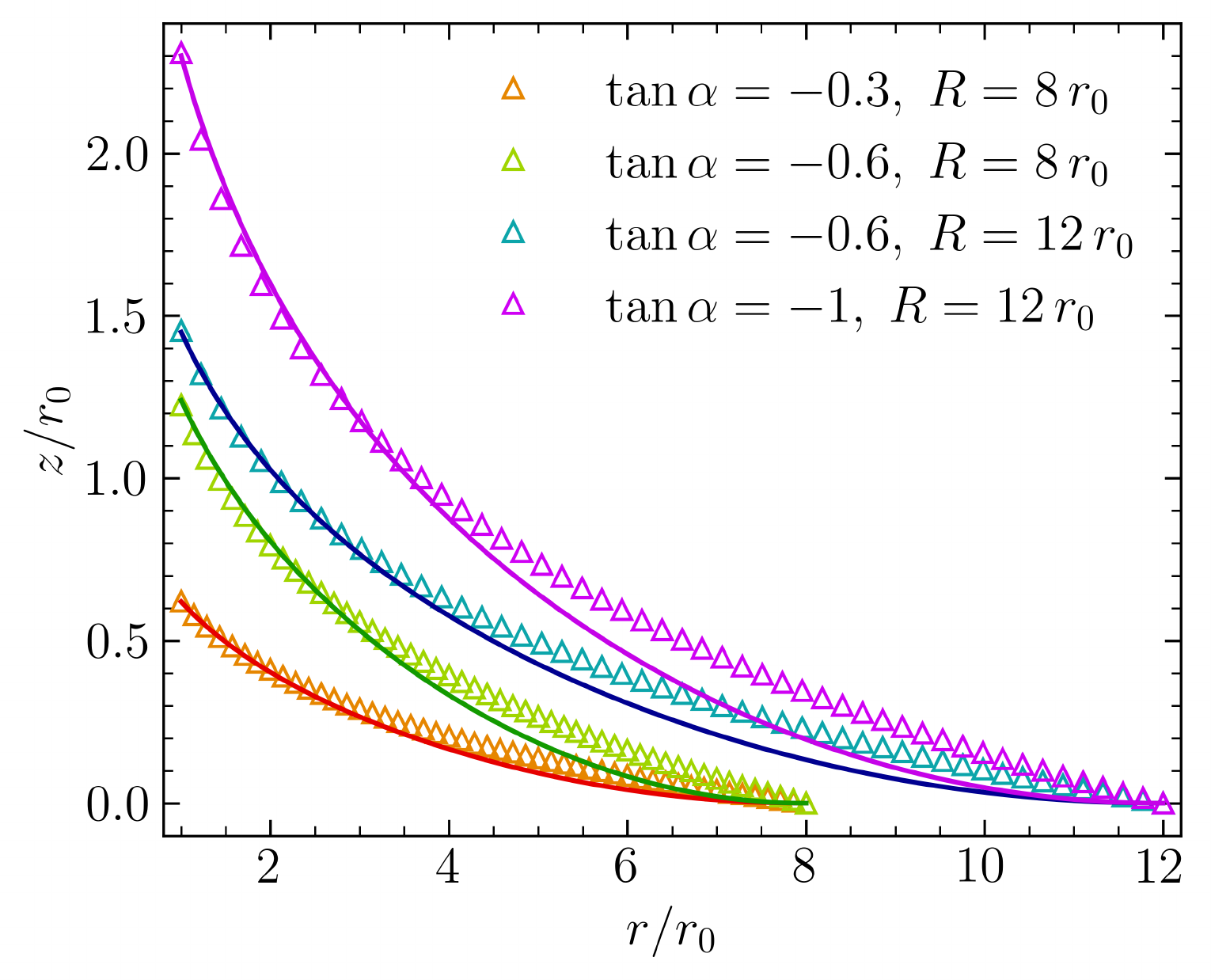}
    \caption{
        Comparison between the  solution with zero mean curvature \crefs{eq_zero_curvature_solution} and the numerically exact, \acf{fe} solution, in the large-deformation regime. The \ac{fe} solution (thin black curves) has been obtained with  the \acf{irene} \cite{worthmullerIRENEFluIdLayeR2025}, and the zero-mean-curvature solution is shown as thick colored curves. Both solutions are shown for  different values of the membrane radius $R$ and contact angle $\alpha$.
    }
    \label{fig:comparison-zero-curvature}
\end{figure}

Although this solution is only valid for zero mean curvature, it constitutes an approximation of the exact solution in the general case where curvature is not zero. However, it allows for explicit, simple expressions of the membrane shape and height, which may be useful on a qualitative level. In particular, here we infer a relation between the contact  angle $\alpha$ and the corresponding spontaneous displacement of the protein. Indeed, substituting \cref{eq_zero_curvature} in \cref{eq_sub_omega}, and integrating, we obtain that the corresponding profile of the membrane is:
\begin{equation}
    \label{eq_z_zero_curvature}
    z(r) = -C_1 \ln \Big(r -\sqrt{r^2-C_1^2}\Big)  + C_2.
\end{equation}
\revision{\Cref{eq_z_zero_curvature} is the catenoid, a well known zero-curvature profile \cite{powersFluidmembraneTethersMinimal2002}.}
The integration constants $C_1$ and $C_2$ can be determined from the  \acp{bc} \crefs{linearized_BC_1,linearized_BC_3}, leading to:
\begin{equation}\label{eq_zero_curvature_solution}
    z(r) = - r_0 \sin{\alpha} \ln\frac{r - \sqrt{r^2-r_0^2\sin^2{\alpha}}}{R -\sqrt{R^2-r_0^2\sin^2{\alpha}}}.
\end{equation}
The displacement of the protein is then given by:
\begin{equation}\label{eq_zero_curvature_1}
    h = - r_0 \sin{\alpha} \ln\frac{r_0 -r_0 \sqrt{1-\sin^2{\alpha}}}{R -\sqrt{R^2-r_0^2\sin^2{\alpha}}},
\end{equation}
For $R \gg r_0$, \cref{eq_zero_curvature_1} becomes
\begin{equation}
    \label{eq_displacement_zero_curvature}
    h \simeq - r_0\sin{\alpha}\left[\ln \frac{r_0}{R}  + \ln\left(1 - \cos{\alpha}\right)\right]
\end{equation}
The comparison between the numerical solution obtained with \ac{irene} and the zero-mean-curvature solution \crefs{eq_zero_curvature_solution} is shown in \cref{fig:comparison-zero-curvature}. The Figure shows that the zero-mean-curvature solution yields a reasonable approximation for the numerically exact solution, see \cref{sec:disc}.

\subsubsection{Forces}\label{sec_forces}

Another important feature in the interaction between the protein and the membrane is the force exerted by the membrane on the \ac{prin} boundary.
The expression for this force is derived in
\cref{app:variation-of-HC}, and its tangential and normal components read
\begin{align}
    \label{eq_f_tan}
    {\bm f}_\perp     & =2\kappa n^i\nabla_iH                    \newlinenn
    \label{eq_f_norm}
                      & =                  -2\kappa \pder{H}{r},            \\
    {\bm f}_\parallel & =-n^i(2\kappa H^2+\sigma){\bm e}_i
\end{align}

\begin{figure*}[t]
    \includegraphics[width=1\textwidth]{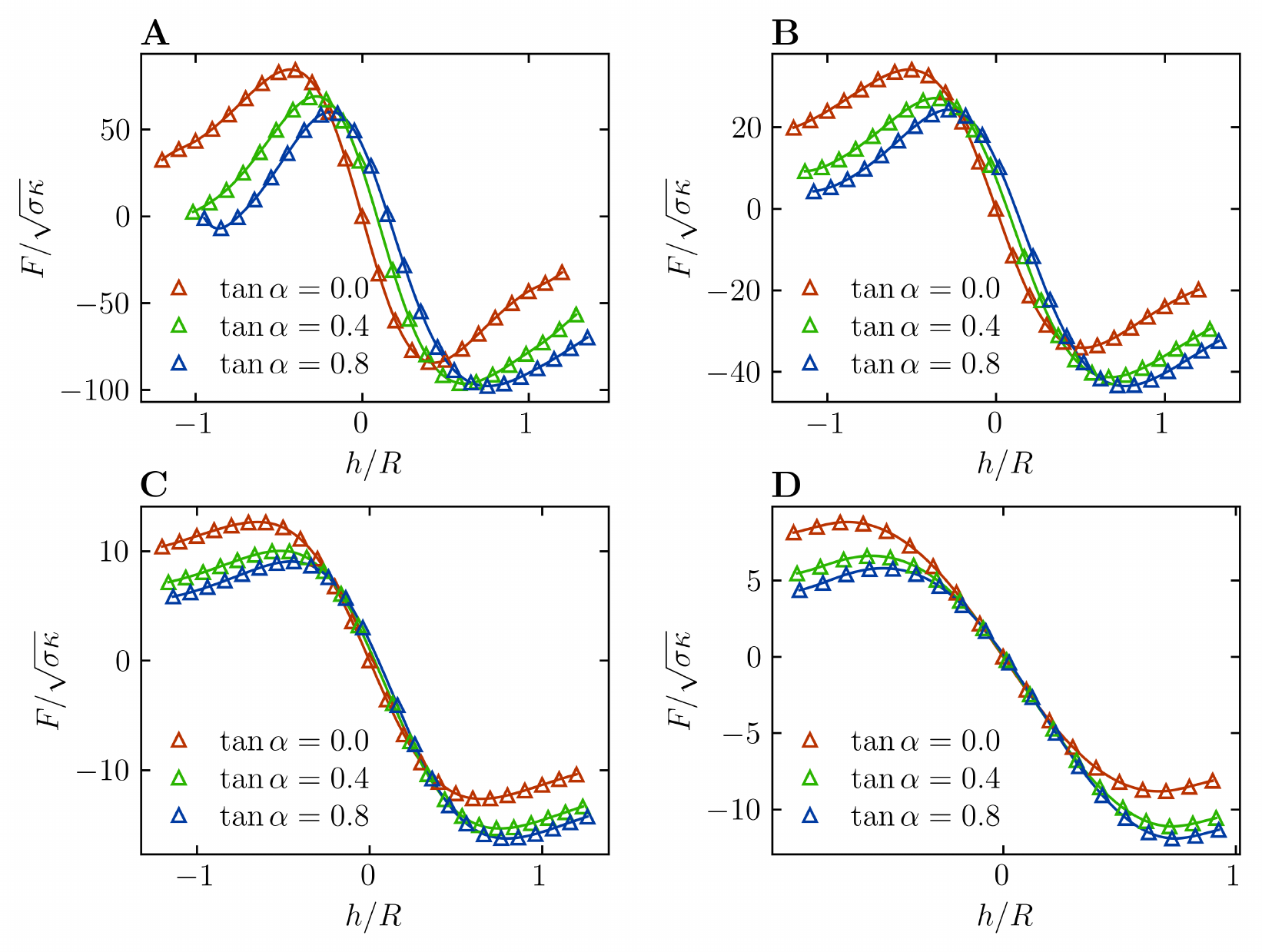}
    \caption{
        Force exerted by the membrane on a \acl{prin} along the \(z\)-axis, in the large-deformation regime. The force has been computed numerically by  using the \acf{irene} \cite{worthmullerIRENEFluIdLayeR2025}. Different panels show the force as a function of the membrane vertical displacement $h$,  for different values of the size $R$ of the membrane domain.
        \textbf{A})        $R = 5 \, r_0$, \textbf{B}) $R = 10\, r_0$, \textbf{C}) $R =30\, r_0$, \textbf{D}) $R = 100\, r_0$.}
    \label{fig:figure_forces}
\end{figure*}

In the \ac{ld} regime, the force exerted by the membrane on the protein  has been evaluated numerically with \ac{irene}; results are shown in \cref{fig:figure_forces}.
\revision{The force is found to be non-monotonic as a function of the membrane vertical displacement $h$, and to depend on the size $R$ of the membrane domain. Comparison of the curves in \cref{fig:figure_forces} with the solution found in \cite{derenyiFormationInteractionMembrane2002} show qualitative adherence. The order of magnitude of the force is found to be in the range of $1-10$ pN, which is the same order of magnitude of forces exerted by molecular motors and actin on the membrane \cite{alberts2022molecular}. Moreover, the presence of a maximum followed by a decrease may suggest the onset of  instabilities in the dynamics, and points to nontrivial behavior in tubule formation and retraction.}

\subsubsection{Potential energy in a system with two proteins}
\label{sec_two_proteins}

\begin{figure}
    \centering
    \includegraphics[width=1.05\columnwidth]{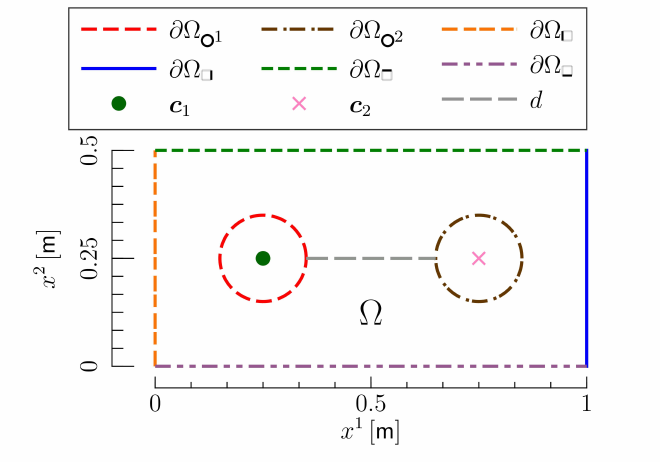}
    \caption{
        Sketch of the domain $\om$ and its boundaries for the steady state of a membrane with two \acp{prin}, in the absence of flows. The domain size is $0 \leq x^1 \leq L$, $0 \leq x^2 \leq h$. The \ac{prin} centers ${\bm c}_1$, ${\bm c}_2$ and the inter-\ac{prin} distance $d$ are also marked.
        \label{fig:domain_two_proteins}
    }
\end{figure}

In what follows, we will focus on the membrane-mediated interaction between the two \acp{prin}. We consider a square domain $\om$ with side $L$, containing two circular holes, which represent the two \acp{prin}, see \cref{fig:domain_two_proteins}.

In this case,  the rotational symmetry  of \cref{fig:BCs-scheme_radial_symmetry} no longer holds. As a result, the \ac{pde} \crefs{eq:stationary_equation_general_form} cannot be reduced to an \ac{ode}, and the analytical treatment of the problem is out of reach; we will thus study the solution numerically.

The centers of the two holes \pomcircone, \pomcirctwo{} are located, respectively, at ${\bm c}_1 = (L/2 - d/2 - r_0, L/2)$ and ${\bm c}_2 = (L/2 + d/2 + r_0, L/2)$. Here,
$d$ is the distance between the two closest points on $\pomcirceqone$ and $\pomcirceqtwo$, and both \acp{prin} have radius $r_0$, see \cref{fig:domain_two_proteins}.
We impose the following \acp{bc}:
\begin{align}
    z            & = 0              \text{ on } \pomsqeq,     \label{two_prot_BC_1}                                         \\
    \nab_i z     & = -\tan \alpha \, \hat{r}_i              \text{ on } {\pom_{\mcirceqcap^{1(2)}}},  \label{two_prot_BC_3} \\
    n^i \nab_i z & = 0              \text{ on } \pomsqeq, \label{two_prot_BC_4}
\end{align}
where $\hat{r}$ is the unit radius in the $x^1\,x^2$ plane relative to the center of each \ac{prin}.

\begin{figure*}
    \centering
    \includegraphics[width=1\textwidth]{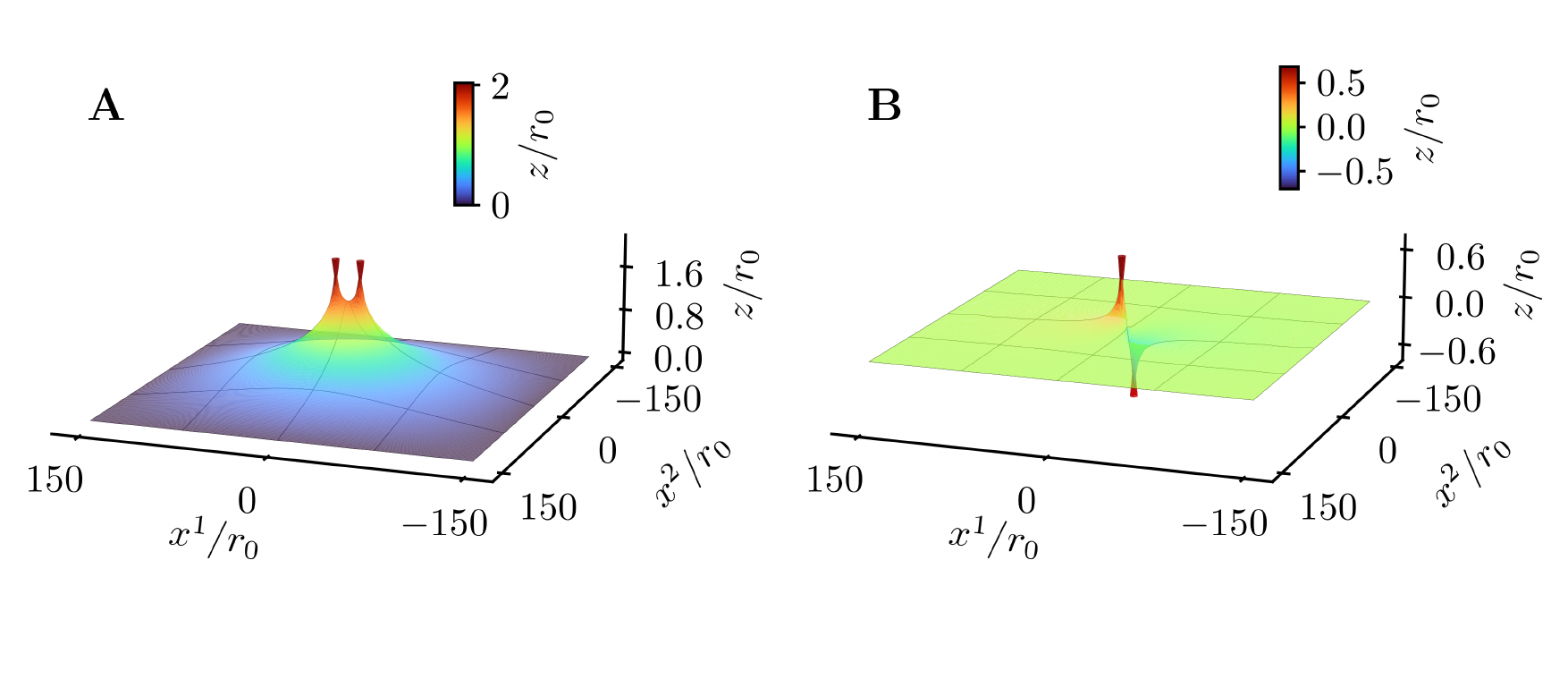}
    \caption{
        Profile of a membrane with two \aclp{prin} in the large-deformation regime, obtained using the \acf{irene} \cite{worthmullerIRENEFluIdLayeR2025}. \textbf{A}) Two \aclp{prin} with contact  angles given by  $\tan \alpha_1 = \tan \alpha_2 = 0.6$. \textbf{B})
        Same as \textbf{A}, with  $\tan \alpha_1 = 0.6$ and $\tan \alpha_2 = -0.6$.
        \label{fig:2proteins}
    }
\end{figure*}

\begin{figure*}
    \centering
    \includegraphics[width=1\textwidth]{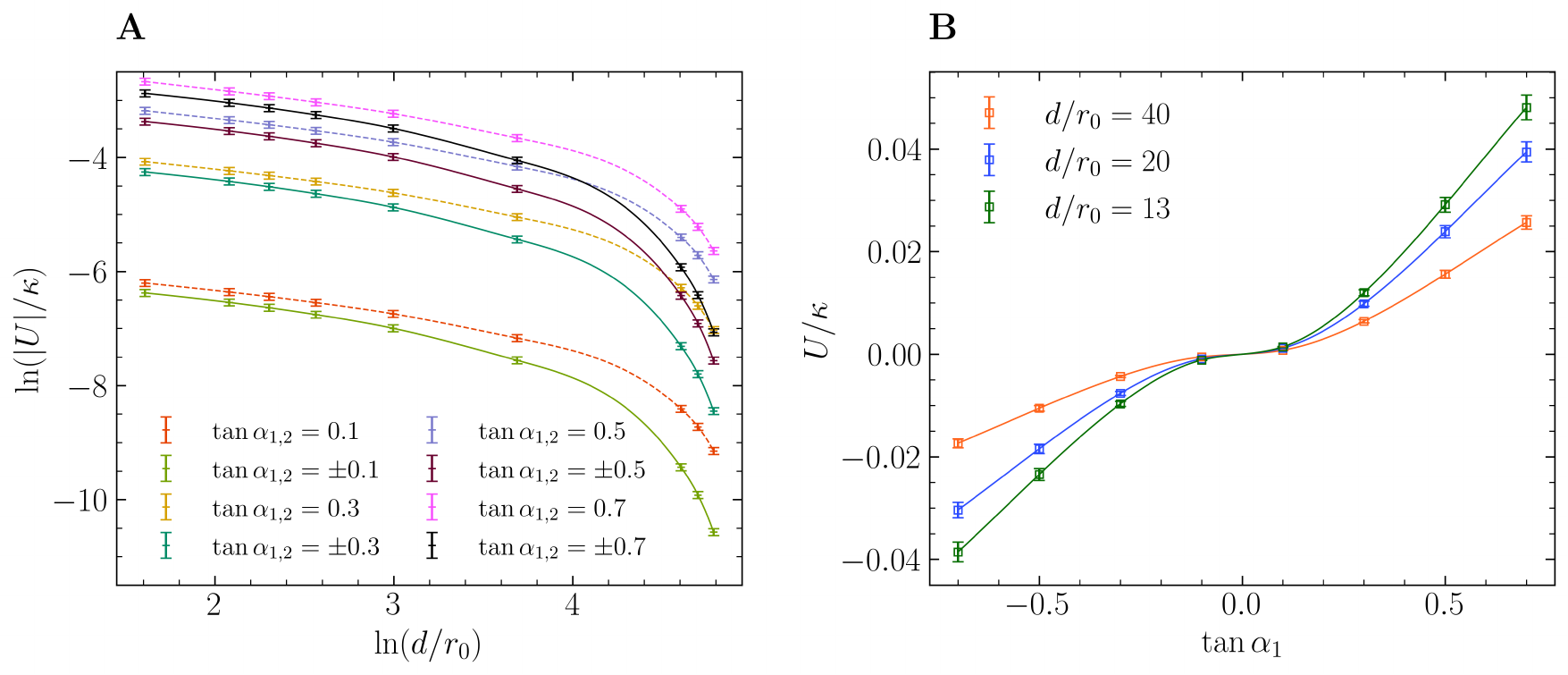}
    \caption{
        Membrane-mediated interaction between two \acfp{prin}. \textbf{A}) Interaction potential \crefs{eq_potential} as a function of the inter-\ac{prin} distance for different values of the contact angles $\alpha_1$ and $\alpha_2$ of the two \acp{prin}. Curves with $\tan \alpha  \gtrsim 0.5$ lie in the large-deformation regime.
        \textbf{B}) Potential as a function of the angle of \ac{prin} $\alpha_1$; the other angle is $\alpha_2 = \left|\alpha_1\right|$.
        \label{fig:2proteins-potential}}
\end{figure*}

\revision{In the case where both proteins impose positive angles, this behavior can be attributed to the additional energy required to deform the membrane in the region between the two \acp{prin} in order to satisfy the imposed angle constraint. This deformation induces significant curvature only over length scales shorter than the crossover length $\ell$.}
On the other hand, when the distance between \acp{prin} is of the order of the protein dimension $r_0$, also the tilt of the phospholipids must be taken into account \cite{fournierCouplingMembraneTiltdifference1998}.

Given that the phospholipid tilt is not currently described by our model, in what follows we will assume the inter-protein distance to be larger than the \ac{prin} radius, $r_0$.
To this end, we numerically compute the energy of the system \cite{helfrichElasticPropertiesLipid1973} over a domain $\Omega$ containing two proteins whose extension is much larger than $d$:
\begin{equation}\label{eq_potential}
    U(d) \equiv \int_{\Omega} \left(\frac{\kappa}{2} H^2 + \sigma \right) \dint{S} - U_\infty,
\end{equation}
where $\dint{S} \equiv \sqrt{\det g} dx^1 dx^2$ is the area element \cite{marchiafavaAppuntiDiGeometria2005}.
In \cref{eq_potential}, $U_\infty$ is the membrane potential  $U(d)$  for $d \rightarrow \infty$.

We considered  \acp{prin} with  contact angles with the same or opposite orientations, see panels A and B, respectively, of \cref{fig:2proteins}.
In addition, \cref{fig:2proteins-potential}A shows the interaction potential as a function of the distance $d$ between the two \acp{prin} for different values of the contact angle.   \Cref{fig:2proteins-potential}B displays the potential as a function of the contact angle of one of the two \acp{prin}, while the other  angle is held fixed. In

the Figure, the term $U_\infty$ in \cref{eq_potential} has been numerically evaluated setting the two \acp{prin} at a distance $d = 100 \, r_0$.
\revision{As one can see from  panel B in \Cref{fig:2proteins-potential}, the energy of the antisymmetric case is lower than that of the symmetric one. For this reason, in situations where both interaction types are allowed on the membrane, the system may exhibit an  antiferromagnetic structure, which may give rise to patterns of alternating proteins.}

% steady state with flows
\subsection{Steady state with flows}
\label{sec_ss_flow}

In this Section, we will extend the analysis of \cref{sec_ss_no_flow} to include the presence of flows in the membrane.

The steady state with flows is described by the \acp{pde}
\begin{align}
    \label{eq_steady_state_flow_1}
    \nab_i v^i  =                & 0,                                                                      \\
    \label{eq_steady_state_flow_2}
    \rho \, v^j \nab_j v^i    =  & \nab^i \sigma + \eta \left( - \nablb v^i   + 2 K v^i   \right) ,        \\
    \nn
    \rho \, v^i  v^j b_{ji}    = & -2\kap \left[  \nab_i \nab^i H  + 2 H (H^2 - K)  \right] + 2 \sigma H + \\
    \label{eq_steady_state_flow_3}
                                 & 2 \eta (\nab^i v^j)b_{ij} ,
\end{align}
where $b$ the second fundamental form \cite{marchiafavaAppuntiDiGeometria2005} \revision{and $\nablb$ the Laplace-Beltrami operator, see \cite{arroyoRelaxationDynamicsFluid2009i,al-izziShearDrivenInstabilitiesMembrane2020} for details}.

\begin{figure}
    \centering
    \includegraphics[width=1.05\columnwidth]{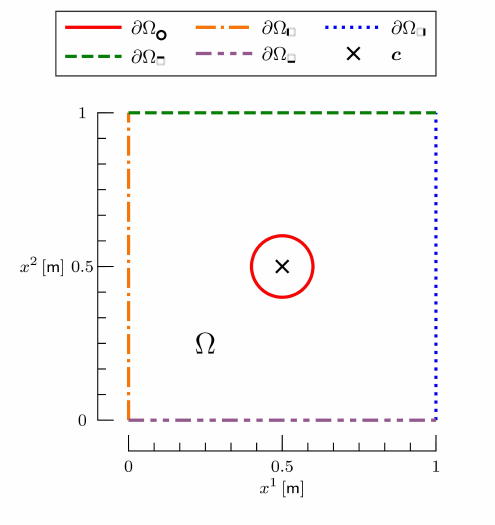}
    \caption{Sketch of the domain $\om$ and its boundaries for the steady state of a membrane with a \acl{prin}, in the presence of flows. The domain size is defined as in \cref{fig:domain_two_proteins},  the \acl{prin} center (black cross) is located at ${\bm c} = (L/2, L/2)$, and its radius is  $r_0$.  }
    \label{fig:BCs-scheme}
\end{figure}

We will consider a geometry given by a square domain with a circular hole, the \ac{prin}, see \cref{fig:BCs-scheme}.
The reference frame is taken to be that of the \ac{prin};  this means that the flow around the \ac{prin}  is studied as if the \ac{prin} were stationary, and the surrounding fluid were moving. Importantly, here we  assume that membrane deformations fall off at a distance from the \ac{prin} comparable to the domain size $L$, or larger.
\revision{In previous studies \cite{arroyoRelaxationDynamicsFluid2009i,danielsCurvatureCorrectionMobility2016}, the Stokes equation was considered. However, in two-dimensional settings this leads to the well-known Stokes paradox, namely that solutions do not decay at infinity \cite{saffmanBrownianMotionBiological1975,saffmanBrownianMotionThin1976}. This issue is typically resolved by introducing regularization mechanisms, most commonly through a coupling with a surrounding fluid.

    In the present case, we retain the convective term, and neglect the effect of the external fluid. The latter assumption is justified as long as the characteristic length scale of flow-induced deformations, $\lambda$, remains smaller than the Saffman--Delbrück length $\lambdasd$, which represents the scale at which viscous stresses within the membrane become comparable to those exerted by the surrounding fluid. The Saffman--Delbrück length is given by $\lambdasd = \eta / \etathreed$, where $\etathreed$ is the viscosity of the external fluid (taken as $\etathreed = 10^{-3} \, \pas  \sec$). For the parameters considered in this work, $\lambdasd$ is of the order of $10^{-5}\,\met$, corresponding to approximately $10^3 \, r_0$, which is one order of magnitude larger than the computational domain size $L = 100\,  r_0$.}

Given that the square boundary removes the rotational symmetry, the \acp{pde} \crefs{eq_steady_state_flow_1,eq_steady_state_flow_2,eq_steady_state_flow_3} can no longer be reduced to a set of \acp{ode}, and a full numerical solution is needed. We  numerically solve \cref{eq_steady_state_flow_1,eq_steady_state_flow_2,eq_steady_state_flow_3} by imposing the following \acp{bc}, see \cref{fig:BCs-scheme}:

\begin{align}
    \label{bc_flow_3}
    v^1          & = v_0 \text{ on } \pomineq,                        \\
    \label{bc_flow_4}
    v^2          & = 0 \text{ on } \pomineq,                          \\
    \label{bc_flow_1}
    v^1          & = 0 \text{ on } \pomcirceq,                        \\
    \label{bc_flow_2}
    v^2          & = 0 \text{ on } \pomcirceq,                        \\
    \label{bc_flow_5}
    \sigma       & = \sigma_0 \text{ on } \pomouteq,                  \\
    \label{bc_flow_5b}
    n_i \Pi^{i1} & = 0 \text{ on } \pomouteq,                         \\
    \label{bc_flow_6}
    n^i v_i      & = 0 \text{ on } \pomweq,                           \\
    \label{bc_flow_7}
    \nab_i z     & = t_\mcirceqcap \hat{r}_i  \text{ on } \pomcirceq, \\
    \label{bc_flow_8}
    n^i \nab_i z & = 0 \text{ on } \pomsqeq,                          \\
    \label{bc_flow_9}
    z            & = 0 \text{ on }  \pomsqeq.
\end{align}
In \cref{bc_flow_5b}, $\Pi_{ij} \equiv - \sigma g_{ij} - \eta (\nab_i v_j + \nab_j v_i)$ is the membrane stress tensor \cite{al-izziShearDrivenInstabilitiesMembrane2020,arroyoRelaxationDynamicsFluid2009i}. In \cref{bc_flow_7}, $\hat{r}$ is the unit radius in the $x^1\,x^2$ plane relative to the \ac{prin} center $\bm c$, and we have set
\be
\pomsqeq \equiv \pomineq \cup \pomouteq \cup \pomweq.
\ee

The no-slip \acp{bc} \crefs{bc_flow_1,bc_flow_2}   reflect the assumption that phospholipids in direct contact with the protein are immobilized due to binding interactions \cite{landauFluidMechanics1987}. \Cref{bc_flow_3,bc_flow_4} represent flow of membrane elements  from \pomin, where the velocity field is assumed to be uniform and unperturbed by the \ac{prin}.  \Cref{bc_flow_5} reflects the hypothesis that \pomout{} is far enough downstream of the protein that $\sigma$  matches the unperturbed,  intrinsic tension of the membrane, $\sigma_0$. \Cref{bc_flow_6}  enforces the presence of two `walls' on the two sides of the boundary, through which membrane elements cannot flow.  \Cref{bc_flow_7} enforces a fixed contact angle at the \ac{prin}, reflecting  binding between membrane phospholipids and the \ac{prin}.
\Cref{bc_flow_7} imposes that the membrane is flat at the outer boundary, according to the definition above of the length scale $L$.  Finally, \cref{bc_flow_9} fixes the membrane height at the external boundary: This \ac{bc} sets the reference height for the membrane profile, and it also reflects the assumption that membrane deformations decay at distances from the \ac{prin} smaller than or equal to $L$.

\begin{figure*}[t]
    \centering

    \includegraphics[width=1\textwidth]{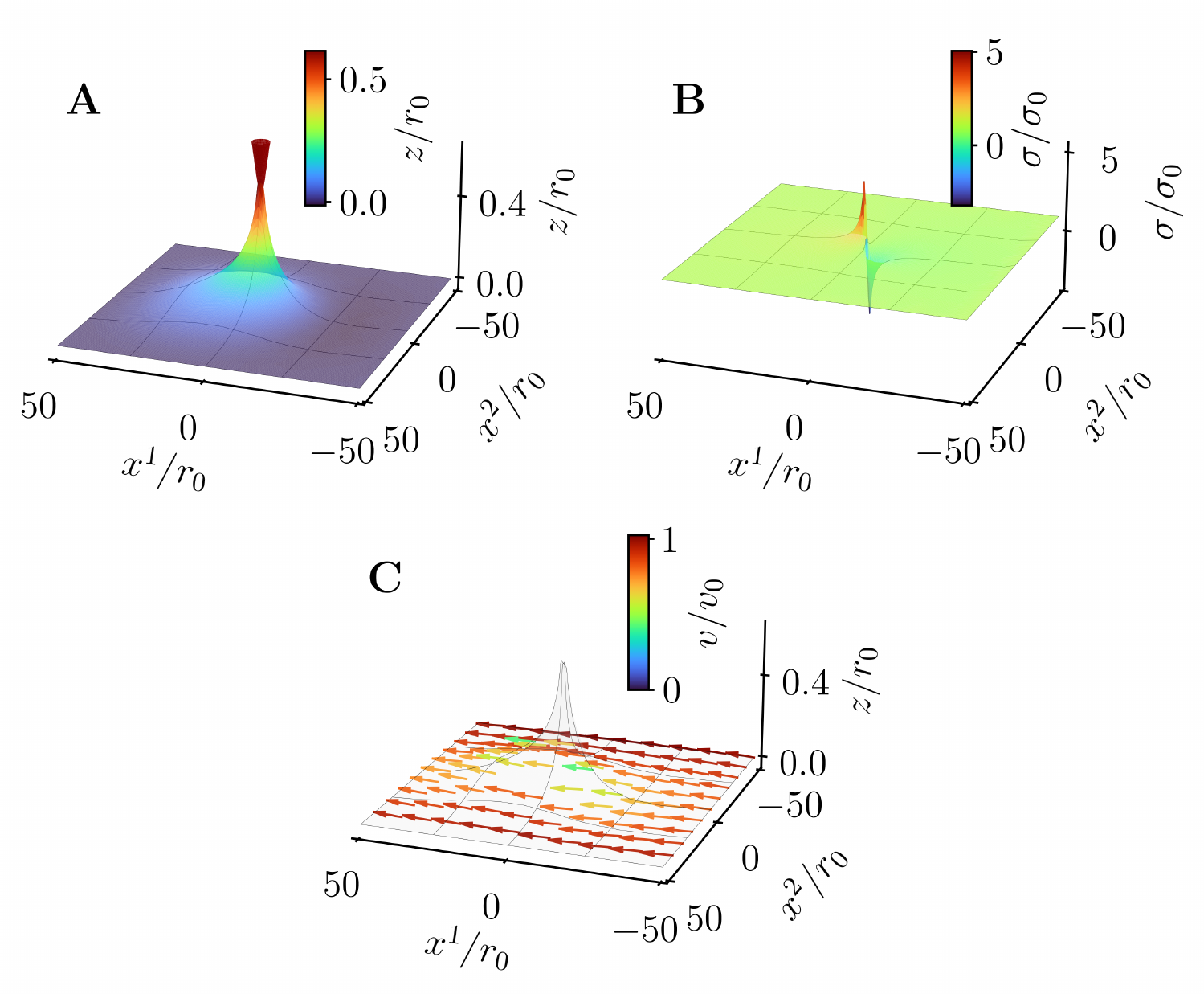}
    \caption{Steady state with flows for a inflow velocity equal to the threshold value,
        $v_0 = \vthreshold$. In \textbf{A}) Membrane shape. \textbf{B}) Membrane tension. \textbf{C}) Membrane velocity, where the velocity direction is shown by the arrows, and its norm by the color bar.}
    \label{fig:fig_dynamic}
\end{figure*}

In \cref{fig:fig_dynamic} we display the resulting numerical solution obtained  by  using \ac{irene}.
The parameters used in \cref{fig:fig_dynamic} are $v_0  = \vthreshold$, where
\be\label{eq_def_v}
\vthreshold \equiv \frac{\kappa}{L \eta},
\ee
$L = 10^2 \, r_0$, $\sigma_0$ is given in \cref{tab:physical_parameters}, and
$t_\mcirceqcap = -0.3$.

\begin{figure*}[t]
    \centering
    \includegraphics[width=1\textwidth]{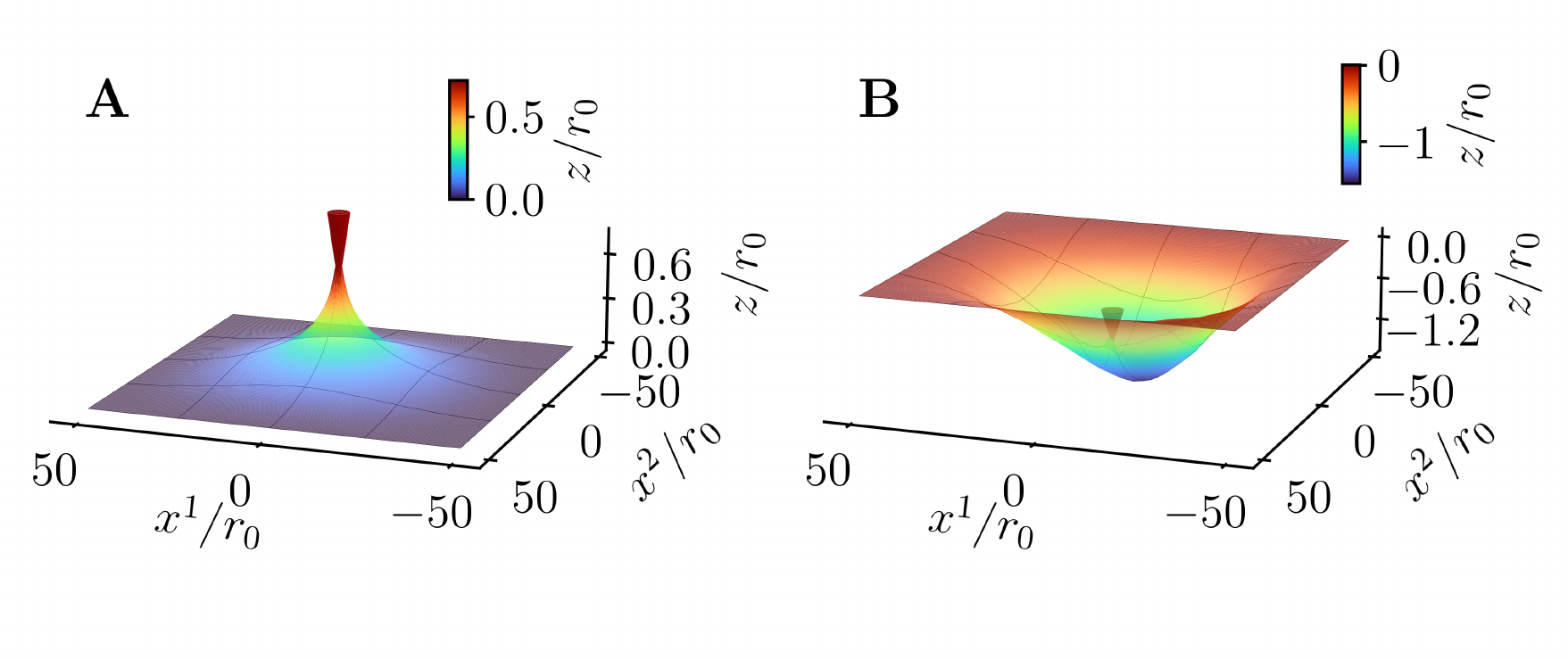}
    \caption{Steady state with flows: membrane shape for different values of the influx velocity. Panels \textbf{A} and \textbf{B} display the membrane shape for
        $v_0$ below and above the threshold velocity,  $v_0= 0.1 \, \vthreshold$ and $v_0= 5\,  \vthreshold$, respectively.}
    \label{fig:dynamical-regimes}
\end{figure*}

\revision{The value of the characteristic velocity, $\vthreshold$, results from an adimensional number  in   \cref{eq_steady_state_flow_3}, i.e.,  the Scriven-Love number \cite{sahuGeometryDynamicsLipid2020,scrivenDynamicsFluidInterface1960,love1944a}
    \be
    \scrivenlovenum = \frac{\eta v L}{\kappa}.
    \ee
    The value of $\vthreshold$ directly follows from setting $\scrivenlovenum=1$.}
As shown in \cref{fig:dynamical-regimes}, the characteristic velocity $\vthreshold$ separates two different physical regimes:
\begin{itemize}
    \item $v_0 < \vthreshold$: The flow has no visible effect on the membrane shape, see \cref{fig:dynamical-regimes}A.
    \item $v_0 > \vthreshold$: A membrane deformation induced by the flow appears, see \cref{fig:dynamical-regimes}B.
\end{itemize}

\revision{From the physical point of view}, the existence of these regimes can be explained as follows.
The characteristic length, or wavelength, $\lambda$, of a flow-induced deformation for a given     flow velocity $v$, can be estimated by comparing the orders of magnitude of viscous and  bending-rigidity effects.
In fact, for a deformation of wavelength $~\lambda$ to appear, the energy cost due to viscous effects, is larger than the bending rigidity cost if $ \lambda \eta v \gtrsim \kappa$.
As a result, if $v \gtrsim v_\ast$, a flow-induced deformation with wavelength $\sim \kap/(v \eta) < L$ appears. On the other hand, if $v \lesssim v_\ast$, the flow-induced deformation wavelength would be larger than the domain size $L$: Given that we assumed that no membrane deformations propagate to a distance equal or larger than $L$ from the \ac{prin}, see \cref{bc_flow_9,bc_flow_8}, for these velocities no deformation can appear, see \cref{fig:dynamical-regimes}.

% discussion 
\section{Discussion}\label{sec:disc}

\acresetall

We have studied, by combining analytical and numerical methods, the deformations in a cell membrane induced by \acp{prin} in the \ac{ld} regime. Our numerical analysis allows for going beyond current perturbative approaches \cite{goulianLongRangeForcesHeterogeneous1993,weiklInteractionConicalMembrane1998,dommersnesLongrangeElasticForces1998,kimCurvatureMediatedInteractionsMembrane1998,marchenkoElasticInteractionPoint2002,bartoloElasticInteractionHard2003,muellerGeometrySurfaceMediated2005}, which cannot describe the \ac{ld} regime.

%sign

We first focused on the steady state of a lipid membrane with a single \ac{prin} with \revision{axisymmetry}, where the membrane is pinned on an outer circle with radius $R$ and the \ac{prin} is located at the center. We showed the limitations of perturbative solutions by comparing them to numerically exact, \ac{fe} solutions \cite{worthmullerIRENEFluIdLayeR2025},  for large protein displacements or large \ac{prin} contact angles, see \cref{fig:comparison-sections}. We developed an analytical solution for the shape equations in the \ac{ld} regime, which assumes that the mean membrane curvature vanishes. Although this solution is only approximate, it yields a sensible, analytical expression of the membrane shape, see \cref{fig:comparison-zero-curvature}.
According to this solution, a \ac{prin}  with a fixed contact angle has a spontaneous displacement that scales, for large $R$, as $\sim \ln R$, see \cref{eq_zero_curvature_1}. Moreover, for a straight contact angle $\alpha = \pi/2$, the spontaneous displacement remains finite, see \cref{eq_zero_curvature_1}.
We show that the vertical component of the force exerted by the membrane on the \ac{prin}---the only nonzero component due to rotational symmetry---displays a non-monotonic dependence on the  vertical displacement $h$ of the \ac{prin}. This behavior is   analogous to the one obtained when a point-like force is applied to a membrane \cite{derenyiFormationInteractionMembrane2002},  demonstrating that the detailed size of the  \ac{prin} does not affect qualitatively the force profile.

We then analyzed  the force exerted by the membrane on the \ac{prin}, in the \ac{ld} regime. The force displays a non-monotonic dependence on the vertical displacement of the \ac{prin}, reminiscent of that of a point-like force applied to a membrane \cite{derenyiFormationInteractionMembrane2002}. We  assesses how this non-monotonic behavior is affected by the \ac{prin} contact angle, as well as by the size of the membrane domain, see \cref{fig:figure_forces}.

In addition, we numerically studied the membrane-mediated interaction between two \acp{prin} in the \ac{ld} regime.
As shown in \cref{fig:2proteins-potential}A, for the values of the inter-protein distance $d$ that we considered,  the potential does not display a power-law behavior. Instead, the potential shape shows a sub-power-law decay, i.e., it decreases with $d$ slower than a power law. This nontrivial structure is due to the complex medium which conveys the interaction---the lipid membrane \cite{johannesClusteringMembranesFluctuations2018}. Finally, \cref{fig:2proteins-potential}B shows that the potential is an odd function of the \ac{prin} orientation, and that two \acp{prin} with the same orientation yield a larger energy than two \acp{prin} with opposite orientations. This behavior confirms the interpretation \cite{weiklInteractionConicalMembrane1998} that the angle imposed by each protein can be thought of as a `charge', i.e., \acp{prin} with the same and opposite orientations repel and attract each other, respectively.

In the presence of membrane flows, we identified the emergence of a characteristic velocity $v_\ast$, of the order of a few microns per second, which reflects the competition between
bending and viscous forces. At low velocities, membrane flows have negligible impact on membrane shape. On the other hand, at higher velocities, flow-induced deformations appear. We provided a theoretical estimate for the critical velocity at which this deformation arises, which explains their characteristic length and aligns well with numerical results. The presence of this characteristic velocity scale suggests that \acp{prin} motion may significantly influence long-range membrane deformations and thus, for instance, membrane-mediated interactions.

An example of how flow-induced membrane deformations can influence interactions between \acp{prin} is given by protein clustering on \acp{guv}.
In this regard, we  consider an example involving \ac{br}---the best understood ion-transport protein
\cite{hauptsCLOSINGBACTERIORHODOPSINProgress1999}.
Multiple \ac{br} molecules have been incorporated and diffuse   \cite{kahyaReconstitutionMembraneProteins2001} on the membrane of a \ac{guv}  \cite{waldeGiantVesiclesPreparations2010}. A large \ac{guv} with a $\sim 100\, \mic$ diameter populated with $\sim 10^4$ \acp{br}, yields an average inter-protein distance $d \sim 2 \, \mic$. Given that the  \ac{br} lateral diffusion coefficient of $\sim 1.2 \, \mic^2/\sec$, the typical values of \ac{br} lateral diffusion velocity is $v \sim 2 \, \mic/\sec$. According to the analysis of \cref{sec_ss_flow}, this implies a flow-induced membrane deformation with wavelength $\lambda \sim \kap/(\eta v) \sim 1.5 \, \mic$, comparable with the inter-protein spacing. Given that this wavelength is comparable to the spacing between \acp{br},  flow-induced membrane  deformations may imply  a potential, significant effect  on \ac{br} interactions and patterning on the \ac{guv} surface.\\

Future work could leverage our results to study \ac{prin} clustering in membranes.
In fact, our analysis may be extended to study systems with more than two \acp{prin}, examining how collective effects influence membrane deformation and forces, both with and without membrane flows. This analysis can also provide novel insights on \ac{prin} clusters \cite{radhakrishnanMathematicalSimulationMembrane2012,johannesClusteringMembranesFluctuations2018}, as well as on the effect of membrane-mediated forces on the cluster assembly mechanisms and scaling laws. Finally, a full stability analysis of the governing equations may reveal the existence of flow-driven instabilities at larger velocities, with implications for membrane trafficking \cite{bethaniSpatialOrganizationTransmembrane2010} and mechanosensitive processes in cells \cite{diCellularMechanotransductionHealth2023}.

\section{Acknowledgments}
We would like to thank P. Bassereau, D. Lacoste, P. Sens and D. W\"{o}rthmuller for useful discussions.

\appendix

\section{Geometrical quantities for the zero-curvature solution}
\label{appendix_zero_curvature}

By substituting \cref{eq_sub_omega} the definitions of the geometrical quantities \cite{desernoNotesDifferentialGeometry2004}, we obtain
\begin{equation}\label{eq_geo_sub_omega_psi}
    \begin{aligned}
        \sqrt{\det g} =     & \frac{r}{\sqrt{\psi^2+1}},                                                 \\
        g^{rr} =            & 1+\psi^2,                                                                  \\
        g^{r\theta} =       & 0,                                                                         \\
        g^{\theta \theta} = & 1,                                                                         \\
        H =                 & \frac{1}{2}\left(\frac{\partial \psi}{\partial r} + \frac{\psi}{r}\right), \\
        K =                 & \frac{\psi}{ r} \pder{\psi}{r}.
    \end{aligned}
\end{equation}

\section{Variation of the Helfrich-Canhan's energy functional}
\label{app:variation-of-HC}

In order to find the line forces acting on the membrane boundary, we  evaluate the variation of the Helfrich energy functional \cite{helfrichElasticPropertiesLipid1973} to linear order under a perturbation of the membrane shape of the form
\be
\delta {\bm X} =   {\bm N} \psi + {\bm e}_i \phi^i,
\ee
where $\bm X$ is the three-dimensional position vector of the membrane surface, $\bm N$ is its normal and ${\bm e}_i$ the tangent vectors relative to the coordinate lines $x^i$ \cite{desernoNotesDifferentialGeometry2004,worthmullerIRENEFluIdLayeR2025,marchiafavaAppuntiDiGeometria2005}.
Note that $\bm n$ is a vector in three-dimensional Euclidean space, which is different from the normal $n^i$ defined on the tangent bundle of the membrane manifold \cite{worthmullerIRENEFluIdLayeR2025}, see \cref{sec_linearized_equation}.
Detailed calculation are presented in \cite{desernoNotesDifferentialGeometry2004}, the final result is:
\begin{align}
   \delta \mathcal{H}[z]
    & = \int_\Omega \dint{S} \, [
         \psi \left( 4\kappa H^3 - 2\sigma H - 4\kappa H K \right)+ \newlinenn
    & \nabla_i \phi^i \left( 2\kappa H^2 + \sigma \right)
         + 2\kappa H \nabla^i \nabla_i \psi
      ].
\end{align}

Multiple integrations by parts lead to
\begin{align}
   \delta \mathcal{H}[z]
   = & \int_\Omega \dint{S} \,
   (4 \kappa H^3 - 2 H \sigma - 4 \kappa H K +                \newlinenn
     & 2 \kappa \psi \nabla_i \nabla^i H) \, \psi  +           \newlinenn
     & \int_{\partial \Omega} \dint{l} \,
   2\kappa [H n^i \nabla_i \psi - \psi n_i \nabla^i H +                  \newlinenn
     & + n_i \phi^i (2 \kappa H^2 + \sigma)].
\end{align}

The terms in the integral over the boundary $\partial \Omega$ that are proportional to $\psi$ and $\phi^i$ represent the inverse of the line forces acting on the membrane boundary, that is \cref{eq_f_tan,eq_f_norm} with exchanged sign.

\end{document}